\documentclass[12pt]{article}

\usepackage{amssymb}
\usepackage{graphicx}

\begin{document}

\title{Multichromatic quantum superpositions in entangled two-photon absorption spectroscopy}

\author{M Wittkop,$^{a}$ Juan M. Marmolejo-Tejada,$^{a}$ Mart\'in A. Mosquera$^{a,*}$\\
{\small Department of Chemistry and Biochemistry}\\
{\small 103 Chemistry and Biochemistry Building, Bozeman, MT-59717, USA}\\
{\small ${}^*$\texttt{martinmosquera@montana.edu}}
}

\maketitle
\begin{abstract}
Quantum information science is driving progress in a vast number of scientific and technological
areas that cover molecular spectroscopy and matter-light interactions in general. In these fields,
the ability to generate quantum mechanically-entangled photons is opening avenues to explore the
interaction of molecules with quantum light. This work considers an alternative way of correlating
photons by including energy superpositions. We study how the multichromatic quantum
superposition, or color superposition of photon-pair states, influences the optical properties of
organic chromophores. This work uses electronic structure calculations based on time-dependent
density functional theory, and a simple modification of the standard entangled two-photon absorption
theory. Our calculations show that it is possible to substantially modify the optical absorption
cross section of molecules, where constructive and destructive interferences are
computed. The quantum interference effects are more pronounced than the constructive ones. These
quantum effects, or related ones, could be observed in quantum spectroscopic experiments where qudit
photon states are generated.
\end{abstract}

\section{Introduction}\label{intro}
Quantum spectroscopy offers tools to elucidate molecular systems and materials that both expand and
complement techniques based on classical light
\cite{eshun2022entangled,ma2021nonlinear,saleh1998entangled}.  This has motivated work that is
constantly demonstrating the significant transformative potential of quantum light to understand
molecular function and open technological opportunities. Such advances are taking place in parallel with scientific and engineering fields
such as photonic quantum computing \cite{pelucchi2022potential,slussarenko2019photonic}, where the
precise and accurate control of correlated photons could bring substantial advantages for diverse,
cutting-edge applications. 

A specific phenomenon that has received widespread attention recently is the absorption and emission of
entangled photon pairs by molecules. Entangled two-photon absorption (ETPA)
\cite{schlawin2018entangled}, for instance, introduces physical correlations that are not possible
in classical two-photon absorption (TPA) spectroscopy \cite{varnavski2020two}. ETPA
commonly relies on the generation of entangled photon-pairs, which usually occurs through the
well-known spontaneous parametric down conversion (SPDC) process \cite{zhang2021spontaneous}. SPDC provides
photon pairs where the polarization of the photons are quantum mechanically correlated, and such
entanglement can be verified in EPR-like (Einstein-Podolsky-Rosen) devices \cite{klyshko1988combine,klyshko1988simple,martinez2022witnessing}. These
photons are emitted within an quantum area, and the photons in each entangled pair are delayed with
respect to one another \cite{shih1994two,sergienko1995experimental}. Such delay is expressed in terms of
an entanglement time, which
endow ETPA techniques with unique properties that to-date continue to be explored by the
community.  Entangled photons can be emitted through other mechanisms such as
molecular pathways \cite{rezai2019polarization,lombardi2021triggered}, quantum dots
\cite{basso2021quantum,huber2018semiconductor}, or semiconducting devices
\cite{kues2017chip,grassani2015micrometer}.

A family of chromophores have been investigated recently: molecules such as Rhodamine 6G
\cite{tabakaev2021energy}, Zinc-TPP \cite{burdick2021enhancing}, and thiophene dendrimers
\cite{harpham2009thiophene}, among other dyes \cite{parzuchowski2021setting}. ETPA absorption of a
molecular unit is quantified commonly through ETPA cross-sections, which have the same units as
classical one-photon cross sections, usually expressed in $\mathrm{cm}^2$ units. Rigorous
experimental efforts suggest that ETPA offers considerable quantum advantage over classical TPA
(CTPA) \cite{parzuchowski2021setting}, especially at low photon (quantum light) fluxes (at extremely
low fluxes ETPA will dominate significantly over CTPA). However, the estimation of entangled cross
sections with very high accuracy is the subject of current efforts
\cite{mikhaylov2022hot,parzuchowski2021setting}.  This is a strong motivation, in our opinion, to
further the understanding of the interaction of molecules and materials with quantum light and
their connection to quantum technologies. One can hypothesize that experimental and theoretical
techniques will continue to advance in these directions, unlocking unexpected and highly beneficial
quantum phenomena in a wide variety of systems.  

Experimental and theoretical studies so far have focused on the interplay between single entangled
photon pairs and molecules, where the frequencies of the photons are assumed to be given. For
example, in degenerate pumping the frequency of both photons are the same. Recently, however, energy
superposition of photons has been achieved for individual photons
\cite{zakka2011quantum,treutlein2016photon,clemmen2016ramsey} and for photon pairs
\cite{kues2017chip}. Energy superpositions give rise to the well-known ``qudit'' states, which
generalize the concept of the qubit.  Also known as color superposition, in this phenomenon, the color of
the photons is undetermined, and each single photon is in a superposition of two (or more) colors.
Even though the interaction between photon qubit (or qudit) states and molecules has not been
reported experimentally so far, their effects can be explored theoretically. We do so in this case
for three chromophores of interest: flavin mononucleotide (FMN)
\cite{villabona2018two,homans2018two}, topotecan (TPT)
\cite{burke1996fluorescence,so2000two,khan2018exploring,di2013spectroscopy}, and lucifer yellow
(LY) \cite{kristoffersen2014testing}. This work studies the molecular absorption of entangled
photon-pairs that also feature multichromatic superpositions and polarization entanglement. These
photon pairs could form qudits of six-fold dimensionality (or eight-fold if four colors are used);
this work, however, focuses on a qubit representation, as specified herein, but a
higher-dimensionality of quantum states is possible. We find that the cross sections in this case
show signatures of constructive and destructive interferences, depending on the location of the
quantum superposition in the Bloch sphere. For a special set of angles in the Bloch sphere, we
notice that the destructive interference can be quite substantial, and for other angles, we
observed that the absorption cross section can be enhanced by close to an order of magnitude,
whereas the quantum interference can lower absorption by up to two orders of magnitude.  These
findings then suggest the phenomenon of color-superposition could be of interest for additional
quantum control of ETPA cross sections and related experiments.


\section{Theory}
This work focuses on the interaction of photon pairs with vertical electronic excited states only.
This is a common approximation that is employed in CTPA spectroscopy because full vibronic
transitions are quite demanding, computationally.

For photon-pairs characterized by two unique frequencies $\omega_1$ and $\omega_2$, entanglement
time $T_{\mathrm{e}}$, and entanglement area $A_{\mathrm{e}}$, the absorption cross-section is given
by \cite{fei1997entanglement}: $ \sigma_{f,0}=(4\pi^3\alpha a_0^5)/(A_{\mathrm{e}} \tau_0 c)\times
g(\omega_{\mathrm{T}}-\Omega_f)|W_{f,0}|^2 $, where $g$ is the line shape function, $\omega_{\mathrm{T}}$ is the sum of the two photon frequencies, $\tau_0$ is the atomic unit
of time ($\tau_0=m_e a_0^2/\hbar$), $\Omega_f$ is the excitation energy for transition from the
ground state to the excited state labeled $f$ (the ground state is labeled as the $0$-th state), 
$a_0$ is the Bohr length, $\alpha$ the fine structure constant, and $c$ the speed of light. The
cross-section then has the units that arise from the term $(4\pi^3\alpha a_0^5)/(A_{\mathrm{e}}
\tau_0 c)$. This result for $\sigma_{f,0}$ can be derived using second perturbation theory and
assuming that the photons are modeled as uniform plane wave packets with fronts that are
spatially separated by a distance of $c T_{\mathrm{e}}$. In terms of random photon detection times,
the time difference between photon detections is then in average the entanglement time,
$T_{\mathrm{e}}$. 

The function $W_{f,0}$ reads: \begin{equation} W_{f,0}(\omega_1, \omega_2,
T_e)=\sqrt{\frac{\omega_1\omega_2}{T_e}}S_{f,0} \end{equation} where $\omega_1$ and $\omega_2$ are
the frequencies of the first and second incoming photons, correspondingly. In the expressed equations, all the frequencies,
dipole moments, and entanglement times are expressed in atomic units (a.u.); this includes $\kappa$
and $\Gamma$. The term $S_{f,0}$ represents the transition function (in atomic units):
\begin{equation} S_{f,0}=\sum_j\frac{(\vec{\mu}_{fj}\cdot \vec{\epsilon}_2)(\vec{\mu}_{j0}\cdot
\vec{\epsilon}_1)}{
\Omega_{j}-\omega_1-\mathrm{i}\kappa}\Big\{1-\exp\big[-\mathrm{i}(\Omega_{j}-\omega_1-\mathrm{i}\kappa)
T_{\mathrm{e}}\big]\Big\} +(1\leftrightarrow 2) \end{equation} where $\kappa$ represents the
inverse of the lifetime of the intermediate virtual state, $\vec{\epsilon}_1$ and
$\vec{\epsilon}_2$ are the polarizations of the first and second photon, respectively. The
transition dipole vector for a transition from ground state to excited state ``$j$'' is denoted as
$\vec{\mu}_{j0}$, whereas $\vec{\mu}_{fj}$ denotes the transition dipole vector for a transition
from the $j$-th excited state into the final state $f$. The lineshape function $g$ is a described
in terms of a Lorentzian profile of the form $g(\Delta \omega)=\pi^{-1}
(\Gamma/2)/[\Delta\omega^2+(\Gamma/2)^2]$.

\begin{figure}
    \centering
    \includegraphics[scale=0.4]{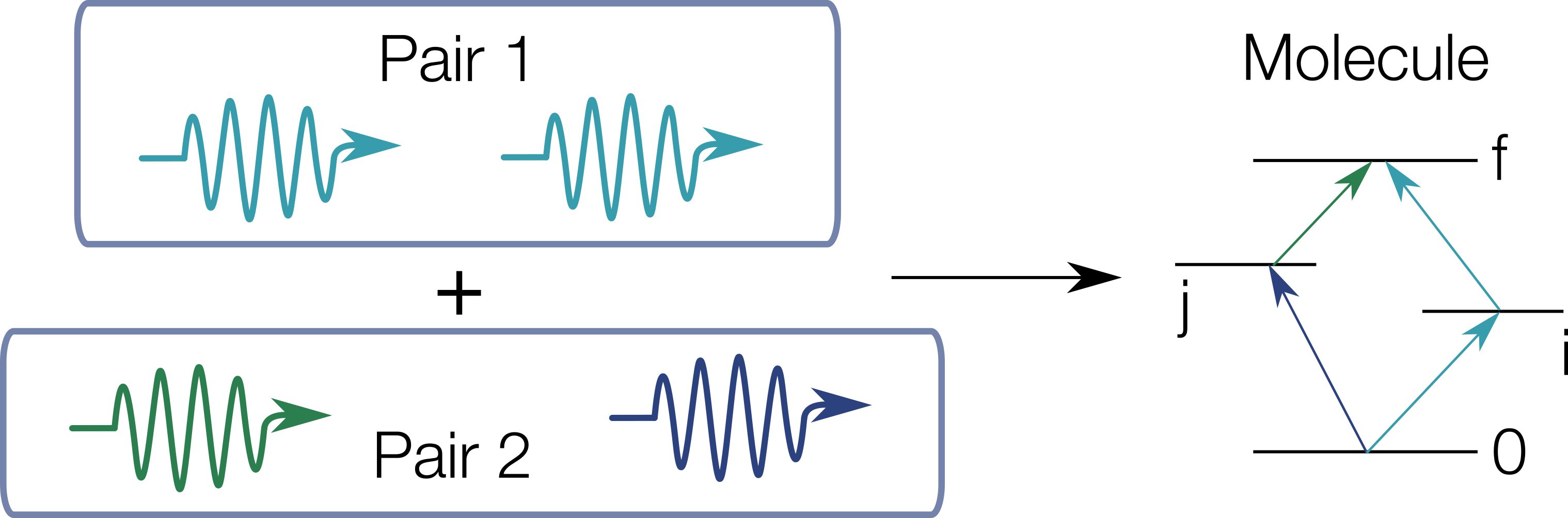}
    \caption{Pictorial representation of the present theoretical model: A two-photon quantum packet
    is directed towards a molecule. The two-photon state is a quantum superposition of two possible
    entangled states, a monochromatic pair, labeled ``MC'', and a bichromatic pair, referred to as
    the ``BC'' pair. The molecule can then be excited through different pathways where different
    intermediate states are involved in the overall two-photon excitation, as suggested in this
    figure.}
    \label{fig:model}
\end{figure}

We now consider the two-photon packet as being in a superposition of two states: ``MC'', in which
the photons are monochromatic ($\omega_1=\omega_2=\omega_{\mathrm{T}}/2$), and ``BC'', in which the
photons are bichromatic, $\omega_1'\neq\omega_2'$; the ``primed'' quantities are assigned to the BC
state. It is assumed that MC and BC photon-pair quantum states have
entanglement times $T_{\mathrm{e}}$ and $T_{\mathrm{e}}'$, respectively, but have assigned the same
area $A_{\mathrm{e}}$. The quantum superimposed two-photon state is thus described by:
\begin{equation}
|\Psi_{\gamma}\rangle =
\cos\Big(\frac{\theta}{2}\Big)|\mathrm{MC}\rangle+
\sin\Big(\frac{\theta}{2}\Big)e^{\mathrm{i}\phi}|\mathrm{BC}\rangle
\end{equation} 
we refer to this state as a ``multichromatic superposition'' (MCS). The MCS is thus controlled by the
Bloch sphere parameters $\theta$ and $\phi$, where $0\le \phi \le 2\pi$ and $0\le \theta\le \pi$.
Because the MCS photon configuration obeys the standard linear superposition of states, the function
$\tilde{W}_{f,0}$ must also transform as: 
\begin{equation}\label{eq_main}
W_{f,0}^{\mathrm{QS}} = \cos\Big(\frac{\theta}{2}\Big)
W_{f,0}^{\mathrm{MC}}+\sin\Big(\frac{\theta}{2}\Big) e^{\mathrm{i}\phi}W_{f,0}^{\mathrm{BC}}
\end{equation}
The amplitude $W_{f,0}^{\mathrm{QS}}$, as a quantum mechanical transition element, is also described in terms of
the Bloch-sphere angles $\theta$ and $\phi$. The $W$ functions are evaluated as
$W^{\mathrm{MC}}_{f,0}=W_{f,0}(\omega_1, \omega_2, T_e)$ and
$W^{\mathrm{BC}}_{f,0}=W_{f,0}(\omega_1', \omega_2', T_e')$. Figure \ref{fig:model} shows a pictorial summary of the theoretical concept explored in this work.

The cross section for a discrete MCS transition of interest is: 
\begin{equation}
\sigma_{0\rightarrow f}^{\mathrm{QS}}=\frac{4\pi^3\alpha a_0^5}{A_e\tau_0 c}
g(0)|W_{f,0}^{\mathrm{QS}}|^2 
\end{equation}
where $\omega_{f,0}=\omega_T$, so $g(0)=2/\pi\Gamma$. The above equation is a relatively simple
extension of the standard formula used in standard ETPA spectroscopy, but it now includes the
possibility of there being color-superposition. This formula supposes, as expected, that dissipation effects
and thereby decoherence in the generation of these MCS photon states are minimal. It is important to notice that perturbation theory
demands energy conservation, as the interaction time is assumed to last for a very long time. For
this reason we have that $\omega_1+\omega_2=\omega_1'+\omega_2'=\omega_{\mathrm{T}}$.

\section{Computational Method}

\begin{figure}
    \centering
    \includegraphics[scale=0.35]{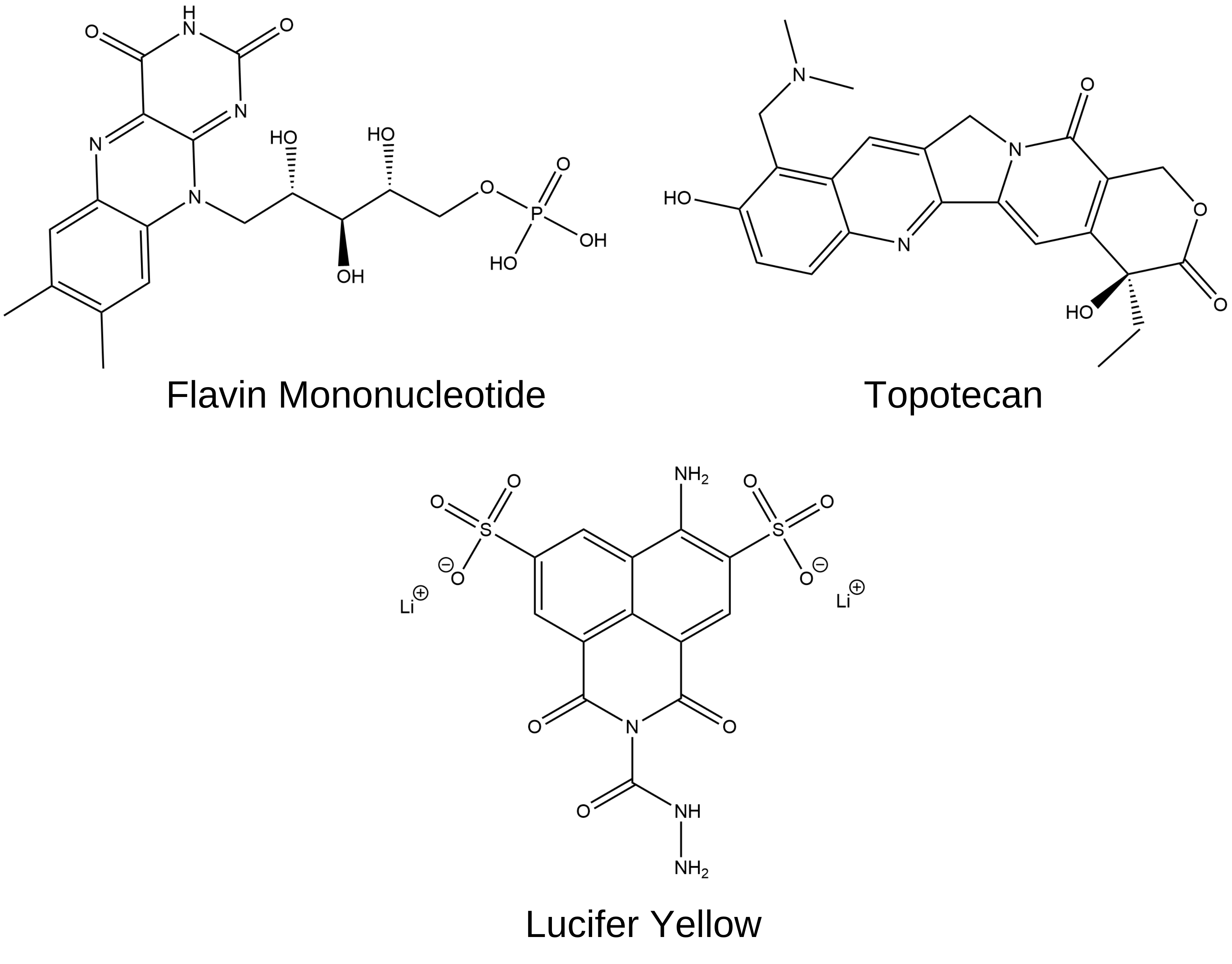}
    \caption{Molecular structures considered in this work: Flavin mononucleotide, topotecan, and lucifer yellow.}
    \label{fig:molecules}
\end{figure}

The transition dipoles are calculated by means of linear-response time-dependent density functional
theory (TDDFT). We use the so-called unrelaxed dipoles, or dipoles that come from ``CIS-like'' 
linear response TDDFT
wavefunctions. This is a practical approximation that works reasonably well, and is computationally
efficient. We extract these dipoles with an in-house code that is based on the quantum chemistry
suite NWChem \cite{nwchem}, version 7.0. We use the standard B3LYP exchange-correlation functional,
and the 6-31G* basis set, which is commonly applied to determine excited-state transitions in the
optical regime.

As discussed before, there is yet a source ETPA enhancement that is not well-understood from a theory
perspective. In Ref. \cite{kang2020efficient}, the use of radiative lifetimes was suggested to obtain
ETPA cross-section values in the range of what has been experimentally observable.  Motivated by the common
practice in TPA theory of using a standard value for the broadening, for the ETPA calculations, we
assume a value of $\Gamma$ that corresponds to $10^{-8}~\mathrm{eV}$, reflecting a final state
lifetime of approximately $1.0~\mathrm{ns}$. For the intermediate state we assume
$\kappa=0.01~\mathrm{eV}$. In all the ETPA calculations the entanglement area is $A_e=1.0\times
10^{-8}~\mathrm{cm}^2$. For the standard ETPA (non-color superposition), we assume $T_{\mathrm{e}}=100~\mathrm{fs}$. This same value, $T_{\mathrm{e}}=100~\mathrm{fs}$, is applied for the 
MCS (color superimposed) ETPA, for the BC and MC quantum states of TPT and LY (Figure \ref{fig:molecules}) we take $T_{\mathrm{e}}=T_{\mathrm{e}}'=100~\mathrm{fs}$, but for FMN we set $T_{\mathrm{e}}=100~\mathrm{fs}$ and
$T_{\mathrm{e}}'=75~\mathrm{fs}$ ($T_{\mathrm{e}}'$ being the entanglement time of the BC 
quantum state).

Classical TPA spectra are computed for comparison as well. For these we take
the standard final state broadening factor, the classical equivalent of $\Gamma$, as
$0.1~\mathrm{eV}$, and intermediate linewidth, the analogue of $\kappa$, as $0.05~\mathrm{eV}$.
We assume that the photons have cross-polarization: so if one photon has horizontal polarization,
the other has vertical polarization. The cross sections are averaged with respect to all possible
molecular orientations \cite{monson1970polarization}.

\section{Results and Discussion}

\begin{figure}
    \centering
    \includegraphics[scale=0.5]{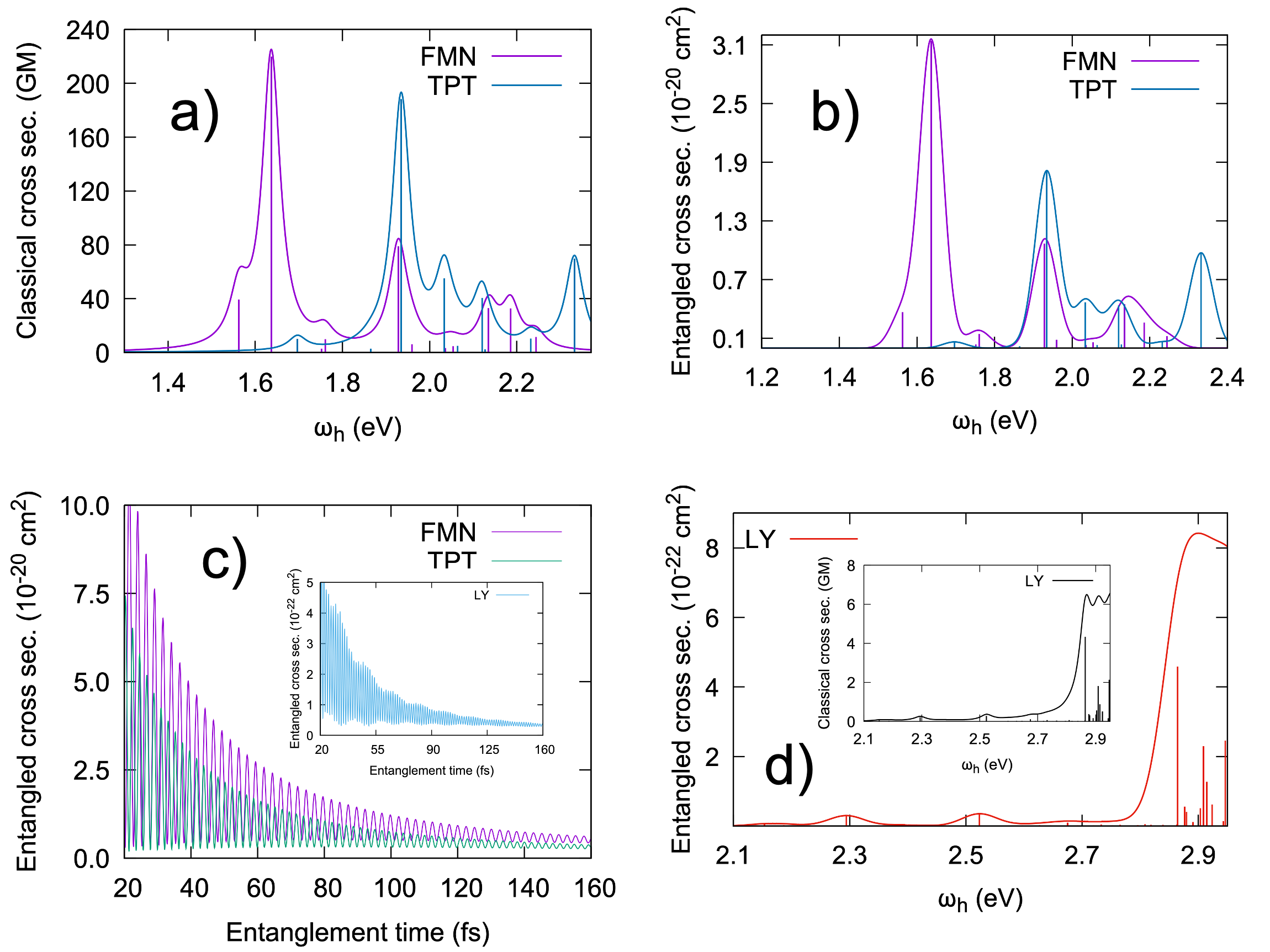}
    \caption{Entangled and classical two-photon absorption cross sections: a) classical TPA profiles
    of FMN and TPT; b), monochromatic ETPA cross section as a function of $\omega_{\mathrm{h}}$ for
    FMN and TPT; c), ETPA cross section vs. entanglement time for FM, TPT, and LY at energies 1.64,
    1.94, and 2.33 eV, respectively. d), ETPA profile of LY, and CTPA cross section (inset).}
    \label{fig:std_etpa}
\end{figure}

To illustrate the proof-of-concept for the effect of color/multichromatic superpositions on the ETPA cross
sections, we first compute the conventional TPA spectrum and the standard ETPA
spectra. Then, we proceed to examine the MCS-ETPA properties of the chromophores selected for our
theoretical study. These chromophores are flavin mononucleotide (FMN), topotecan (TPT), and lucifer
yellow (LY); their molecular drawings are shown in Figure \ref{fig:molecules}. FMN is a biomolelcule produced from riboflavin that forms part of NADH Hydrogenase,
 TPT is used as a chemotherapy drug, and LY is utilized in spectroscopic
studies. The spectra in this work are reported in terms of half the frequency of the total two-photon
frequency. This half-frequency is denoted as $\omega_{\mathrm{h}}=\omega_{\mathrm{T}}/2$.

The theoretical CTPA spectra of FMN, TPT, and LY, are shown in Figure \ref{fig:std_etpa}.a.
and d. (as an inset plot). We note that FMN has a classical TPA cross section of about 220 GM at
around a half-frequency of 1.64 eV (760 nm). As is commonly seen in theoretical TPA spectra, the
classical TPA shows comparable cross-section values at higher $\omega_{\mathrm{h}}$ frequencies. A similar numerical trend is noticed for TPT, where the CTPA cross
section is about 188 GM at 1.94 eV (640 nm), and then it slightly decreases to about 41 GM at 2.12
eV, but it raises again to 70 GM around $\omega_{\mathrm{h}}=2.3$ eV. LY, however, shows a different
profile, as can be seen in Figure \ref{fig:std_etpa}.d, where it is relatively low -  under 1 GM for half-frequencies
between 2.1 and 2.7 eV, where it raises to about 8 GM at 2.9 eV.

\begin{figure}[htp!]
    \centering
    \includegraphics[scale=0.5]{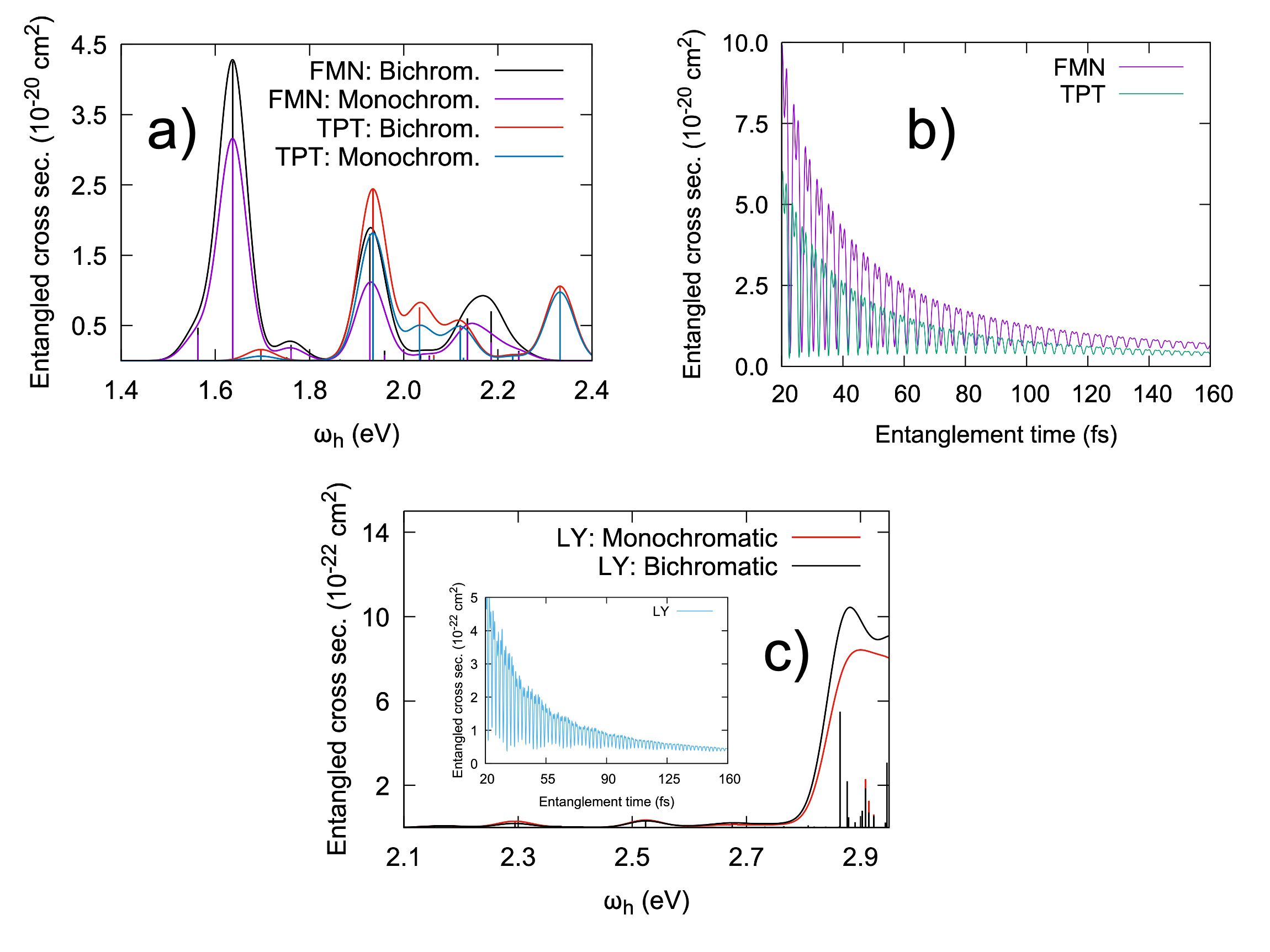}
    \caption{Purely bichromatic ETPA. The frequencies of BC quantum state are $\omega_1'=1/3\times\omega_T$, $\omega_2'=2/3 \times \omega_{T}$: a), BC ETPA spectra of FMN and TPT, and their comparison to the monochromatic counterpart; b), variation of FMN and TPT BC ETPA cross section as a function of the entanglement time, the frequencies $\omega_{\mathrm{h}}$ are the same as in Figure \ref{fig:std_etpa}.c; c), BC ETPA spectrum of LY, inset shows dependency on $T_{\mathrm{e}}$ at 2.9 eV.}
    \label{fig:bichrom_etpa}
\end{figure}

Regarding standard monochromatic ETPA ($\omega_1=\omega_2=\omega_{\mathrm{h}}$), Figure
\ref{fig:std_etpa}.b shows the entangled cross section profiles of FMN and TPT, and Figure
\ref{fig:std_etpa}.d shows that of LY. For these three systems we use
$T_{\mathrm{e}}=100~\mathrm{fs}$. For these systems we see a correlation between the CTPA and ETPA
spectra, however, for TPT such connection between classical and entangled TPA is somewhat less
evident, as there are some changes between relative peak heights. FMN and TPT have ETPA cross
section values in the same order of magnitude, $10^{-20}~\mathrm{cm}^2$. The maximum values of FMN
and TPT occur at the same frequencies as their classical TPA counterparts, with values close to
$3.0\times 10^{-20}~\mathrm{cm}^2$ and $1.8\times 10^{-20}~\mathrm{cm}^2$, respectively. As
mentioned earlier, for the three systems considered we use the same linewidth factors.  The
entanglement time is another variable that affects the magnitude of the ETPA cross sections; the
effect of varying this is displayed in Figure \ref{fig:std_etpa}.c, where we observe, as expected, that
the cross section can vary by one order of magnitude, or more, if the $T_{\mathrm{e}}$ is further
increased beyond the plot time range. For the LY chromophore, we observe theoretical cross section
values lower than those of FMN and TPT, which indicate ETPA activity around 2.3 eV (540 nm) and 2.5
eV (496 nm), and then higher values at 2.9 eV (430 nm).

Figure \ref{fig:bichrom_etpa} displays the standard ETPA spectra under a non-degenerate condition,
where the two photons have different frequencies. In this case we choose $\omega_1'=\omega_T/3$, and
$\omega_2'=2\omega_{\mathrm{T}}/3$ and refer to this as ``bichromatic ETPA''.  The bichromatic ETPA spectrum of
each molecule shows a few differences with respect to the monochromatic ones. These are mainly a
slight enhancement of the ETPA cross section, up to 30 \%.  However, between 2.0 and 2.4 eV, the degenerate-pumping ETPA spectrum of TPT is not improved much by the bichromatic condition. As in the degenerate case, the
cross section at low photon frequency varies by about one order of magnitude for $T_{\mathrm{e}}$ in
the range 20 - 100 fs. Due to the parameter $\kappa$, the ETPA cross-sections in all cases have
their oscillations damped. The oscillatory pattern of the BC ETPA case is slightly different than in
the monochromatic case, but not too significantly.

\begin{figure}
    \centering
    \includegraphics[scale=0.5]{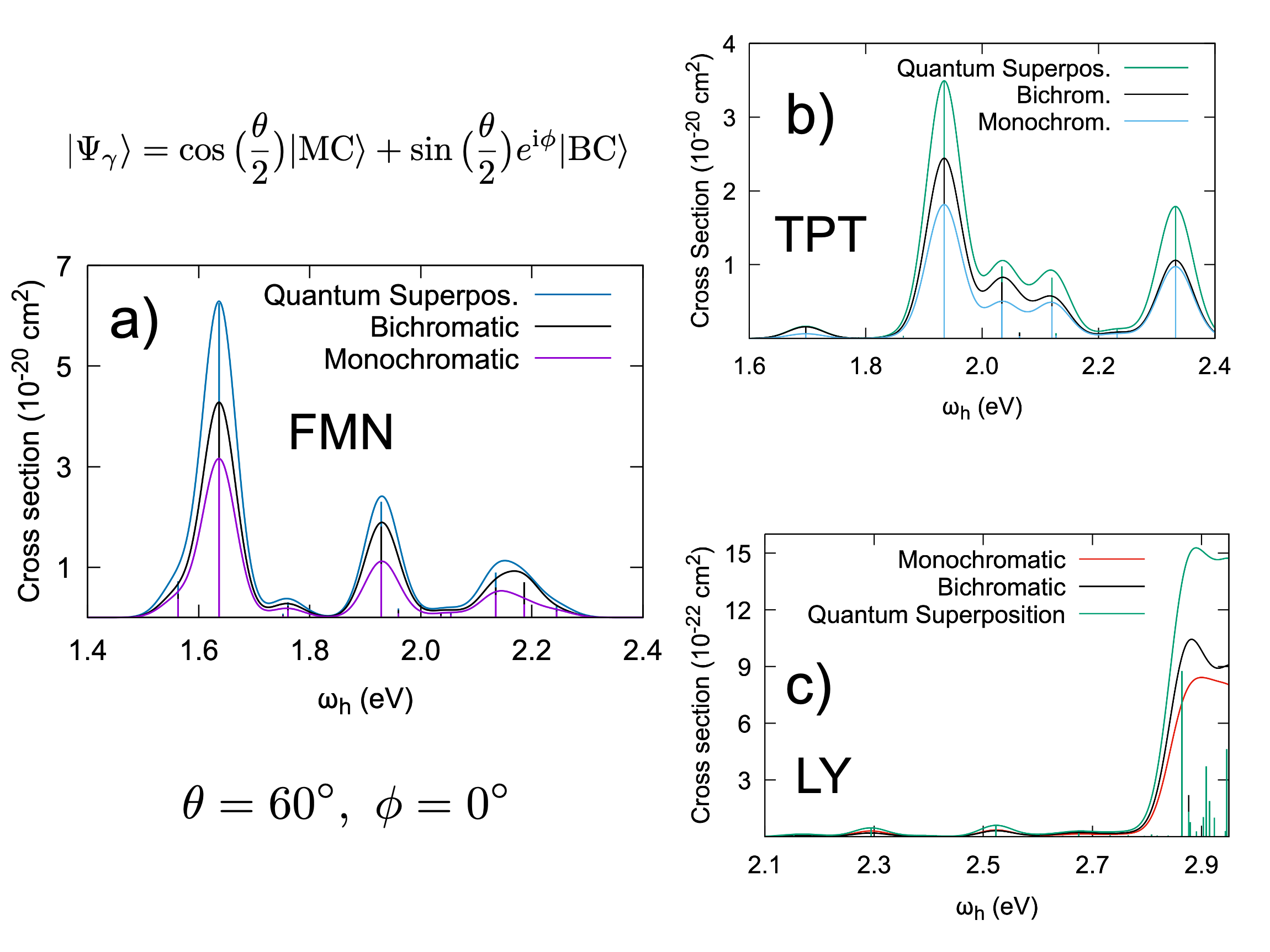}
    \caption{ETPA cross sections based on quantum color superpositions, or quantum MCS. The Bloch sphere parameters are fixed an taken as $\theta=60^{\circ}$ and $\phi=0^{\circ}$. For TPT and LY the entanglement times of the BC and MC states are the same $T_{\mathrm{e}}=T_{\mathrm{e}}'=100~\mathrm{fs}$, but for FMN 
    we study $T_{\mathrm{e}}=100~\mathrm{fs}$ and $T_{\mathrm{e}}'=75~\mathrm{fs}$. a) Shows the standardard quantum MCS ETPA spectra of FMN, b), that of TPT, and c), LY.}
    \label{fig:super_etpa}
\end{figure}

Having determined the difference between the separate MC and BC ETPA cases, we now proceed to examine the
multichromatic quantum superposition effect, as expressed in Equation  (\ref{eq_main}). We
focus on highest intensity transitions, which happen at 1.64, 1.94, and 2.9 eV for FMN, TPT, and LY,
respectively. Figure \ref{fig:super_etpa}.a shows the comparison between the quantum superposition effect and the standard MC and BC ETPA cases where there is no superposition. 
For a polar angle of $\theta=60^{\circ}$, the two-photon state is a quantum mixture of 75 \% the monochromatic state, and 25 \% the bichromatic state, yet we see that the quantum cross section is over 100 \% enhanced with respect to the monochromatic ETPA case, for a frequency of $1.64$ eV. 
A similar enhancement takes place for TPT and LY (Figures \ref{fig:super_etpa}.b and .c, respectively), at 1.94 eV and 2.9 eV, correspondingly. 
This enhancement is also due to setting the phase factor as $\phi=0^{\circ}$. In additional preliminary calculations, we have noted $\phi=0^{\circ}$ leads to quantum {\slshape constructive} effects. 

\begin{figure}
    \centering
    \includegraphics[scale=0.5]{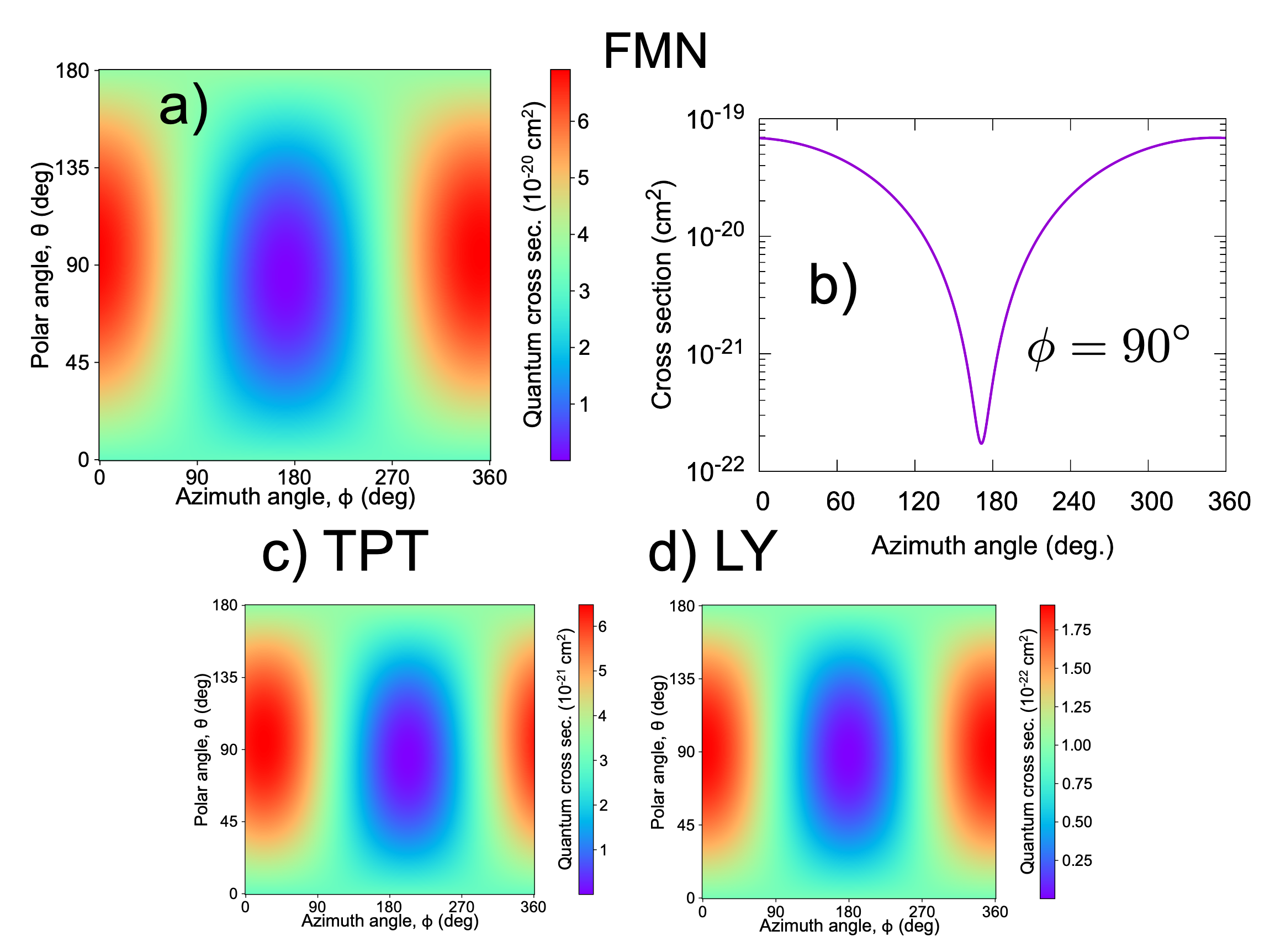}
    \caption{Two-dimensional heat plots for FMN (subfigure a) at $\omega_{\mathrm{h}}=1.64~\mathrm{eV}$, TPT (subfigure c) at 1.94 eV, and LY (subfigure d) at 2.9 eV, and a ``cut'' for $\phi=90^{\circ}$ for the FMN system.}
    \label{fig:2d_etpa}
\end{figure}

For the three molecular systems, FMN, TPT, and LY, Figure \ref{fig:2d_etpa} shows the variation of
cross-section values with respect to the Bloch sphere parameters (again, we examine this at the 
frequencies 1.64 eV, 1.94 eV, and 2.9 eV, correspondingly). The most 
interesting behavior in terms of quantum constructive effects
takes place at the points where theta is between $60^{\circ}$ and $120^{\circ}$, 
while $\phi$ being close to $0^{\circ}$ or $180^{\circ}$. 
There is thus a region of constructive interference, where the enhancement in cross section doubles. 
A higher enhancement is possible for different combinations of entanglement times \cite{giri2022manipulating}, or if the quantum superposition 
has an additional constructive effect that improves the lifetimes of the intermediate and final excited states involved in 
the ETPA quantum process (which are not explored in this work).
On the other hand, a significant {\slshape destructive} quantum interference takes place in a region center around $\theta=90^{\circ}$ and 
$\phi=180^{\circ}$. The cross section value drops by up to two orders of magnitude under the current parameter selection. It drops by two orders of magnitude (from $10^{-20}~\mathrm{cm}^2$ to $10^{-22}~\mathrm{cm}^2$) for FMN,
one order of magnitude for TPT, and two for LY. 
Figure \ref{fig:2d_etpa}.b shows the dependency of $\sigma$ with respect to $\theta$ for $\phi=90^{\circ}$. Clearly, the quantum
superposition engenders behaviors that are not possible under the separate circumstances, which is
to be expected from this type of phenomenon. Therefore, besides present in coherent electronic transfer \cite{bergfield2015harnessing}, destructive quantum interference is also possible in photon absorption.

As is common in quantum coherent phenomena, there are conditions that must be satisfied to observe
quantum superposition effects. On of them is the suppression of decoherence related 
to vibrational degrees of freedom.
These occur during the emission of entangled photons and the absorption of these. It is 
currently a subject of intensive research determining the influence of entangled photons
on the lifetime of excited states. But there are indications that such lifetimes are enhanced 
by entangled photon absorption \cite{ricci2023investigations}. In this work, we assumed fixed linewidth parameters 
in line with values used before \cite{kang2020efficient} to obtain theoretical values consistent 
with experimental measurements. However, as in standard ETPA, the effect of MCS on the lifetimes 
could be a subject of further research as well, as additional quantum correlations could have unexpected 
consequences on electronic lifetimes and other quantum properties.



\section{Conclusion}
In this work we investigated the interaction between entangled photon pairs, that in addition to
having the standard polarization correlation, feature energy superpositions. That is, the
frequency of the photons are undetermined prior to interaction with matter. We referred to this
phenomenon as a multichromatic superposition. The state of the entangled photon pair 
was represented in the
well-known Bloch sphere, or qubit space. In comparison to standard ETPA simulations, we
observed the emergence of constructive and destructive effects in the quantum cross section profiles.
The enhancement (or constructive) of cross section was computed to be improved by nearly a 100 \%,
whereas the desctructive interference could reduce the cross section by around two orders of 
magnitude. Our work then suggests that these types of coherent effects could be added to the toolkits of
quantum control within the context of optical quantum spectroscopy.

\section{Acknowledgments}
The authors kindly thank the MonArk NSF Quantum Foundry supported by the National Science Foundation Q-AMASE-i program under NSF award No. DMR-1906383.


\end{document}